\documentclass{sf2a-conf2012}
\usepackage{graphicx}
\usepackage{hyperref}
\usepackage[]{natbib}  
\usepackage[cyr]{aeguill}
\usepackage{epstopdf}

\def\BibTeX{{\rm B\kern-.05em{\sc i\kern-.025em b}\kern-.08em
    T\kern-.1667em\lower.7ex\hbox{E}\kern-.125emX}}
\bibpunct{(}{)}{;}{a}{}{,}  %%%%%%%%%%%%%  A&A bibliography style
\begin{document}

\TitreGlobal{SF2A 2012}

%%-----------------------------------------------------------------
%%      the top matter
%%

\title{The ELG target selection with the BOSS survey}

\runningtitle{ELG target selection}

\author{S. Escoffier}\address{CPPM, Aix-Marseille Universit\'e, CNRS/IN2P3, Marseille, France}
\author{on behalf of the BOSS Collaboration}
\author{J. Comparat}\address{Aix Marseille Universit\'e, CNRS, LAM (Laboratoire d'Astrophysique de Marseille) UMR 7326, 13388, Marseille, France}
\author{A. Ealet$^1$}
\author{J.-P. Kneib$^2$}
\author{J. Zoubian$^2$}
\author{F. Lamareille}\address{Universit\'e de Toulouse, UPS-OMP, IRAP, Toulouse, France}

%% Keep this line, even if the page will be settled afterwards.
\setcounter{page}{237}

%%-----------------------------------------------------------------

\maketitle

%%-----------------------------------------------------------------
%%        The abstract
%% 
%%  Warning!  within the abstract:
%%  - do not use macros. 
%%  - do not use commands like: \cite, \citet, \citep ... etc.

\begin{abstract}
The Baryon Acoustic Oscillation (BAO) feature in the power spectrum of galaxies can be used as a standard ruler to probe the accelerated expansion of the Universe. In this paper, we study several galaxy selection schemes aiming at building an emission-line galaxy (ELG) sample in the redshift range $0.6 < z < 1.7$, that would be suitable for future BAO studies using the Baryonic Oscillation Spectroscopic Survey (BOSS) spectrograph on the Sloan Digital Sky Survey (SDSS) telescope. We explore two different color selections using both the SDSS and the Canada-France-Hawaii Telescope Legacy Survey (CFHTLS) photometry in the $u, g, r, i$ bands and evaluate their performance for selecting bright ELG. This study confirms the feasibility of massive ELG surveys using the BOSS spectrographs on the SDSS telescope for a BAO detection at redshift $z\sim1$, in particular for the proposed eBOSS experiment.
\end{abstract}

%% Insert the keywords (to appear in the ADS indexing)
%% Keywords must be separated by a comma

%%-----------------------------------------------------------------

\section{Introduction}
%%---------------------

The evidence for the acceleration of the expansion of the Universe \citep{Riess98,Perlmutter99} led to the existence of a new, enigmatic dark energy that opposes the self-attraction of matter. Even though the  $\Lambda$CDM model, which consists of an inflationary cold dark matter model with a cosmological constant, is strongly favored, other interpretations have been suggested to explain cosmic acceleration. Alternatives involve scalar field models as quintessence, modification of the General Relativity itself on cosmological scales or extra dimensions of space-time. There is no doubt that understanding the nature of dark energy is one of the most important puzzles nowadays in cosmology. 

Observational exploration is crucial to provide clues to the energy content of the Universe. One of the leading methods for measuring the expansion history is to use the Baryon Acoustic Oscillations feature (BAO) in the clustering of galaxies as a standard ruler \citep{Seo03}. The BAO refer to the imprint of acoustic waves in the early Universe frozen after the decoupling of baryons and photons. Anticipated as a potential effect of the CMB as early as 1970s \citep{Peebles70,Sunyaev70}, the first convincing BAO detections came in 2005 from the SDSS Data Release 3 (DR3) and the final 2dFGRS samples \citep{Eisenstein05,Cole05}. 
Recently, new BAO detections have been reported. \citet{Beutler11} reported a $4.5\%$ distance measurement at low redshift $z\sim0.1$ with the 6dFGRS sample, and the acoustic scale has been measured in the redshift range $0.16<z<0.47$ by the final SDSS-II sample (DR7) with a distance precision of $2.7\%$ \citep{Percival10,Kazin10}. Steping beyond $z=0.5$, BAO feature has been measured in the SDSS-III Baryon Oscillation Spectroscopic Survey (BOSS;\citep{Dawson12}) (DR9) with a sample of higher redshift luminous red galaxies (LRG) in the redshift range $0.43<z<0.7$ \citep{Anderson12}. Adding to these data, the WiggleZ Dark Energy Survey has quantified BAO by targeting bright emission-line galaxies (ELG) in the range $0.3<z<0.9$ \citep{Blake11}.

In summary, the BAO peak has been detected in several galaxy samples to $z<1$, and there are strong motivations for extending these large-scale structure measurements to higher redshifts. The BAO method measures the cosmic distance in both radial and transverse directions, giving the Hubble parameter $H(z)$ and the angular diameter distance $D_A(z)$, respectively. Beyond the BAO, the full shape of the galaxy power spectrum provides useful information to constrain cosmological parameters and even to test deviations from Einstein's theory of gravity \citep{White09}. Thought to be the cosmological probe the least affected by systematic uncertainties according to the Dark Energy Task Force (DETF;\citep{Albrecht06}), the BAO probe requires to map very large cosmic volumes to achieve a precise distance measurement (as baryons comprise only a small fraction of matter, the amplitude of the BAO peak is small). In addition, resolving the features of BAO along the line-of-sight motivates the need for spectroscopic redshift surveys. The next generation of cosmological surveys (the stage IV facilities as defined in the DETF report) plans to map the high redshift Universe in the range $0.6<z<2$. On the ground, BigBOSS would carry out spectroscopic surveys of 10 million galaxies by targeting LRG to $z=1$ and ELG to $z=1.7$ \citep{Schlegel11}. In space, the ESA's Euclid mission plans to measure redshifts of 50 million strong H$\alpha$ emitters in the redshift range $0.7<z<2$ \citep{Laureijs11}.

Motivated by future BAO surveys as eBOSS\footnote{extendedBOSS is part of a proposed program of post-2014 surveys on the Sloan telescope.} or BigBOSS, this paper deals with a galaxy target selection to identify ELG in the redshift range $0.6<z<1.7$. Section~\ref{escoffier:secELG} describes the color selection using imaging data from the Sloan Digital Sky Survey (SDSS) and the Canada-France-Hawaii Telescope Legacy Survey (CFHTLS). In section~\ref{escoffier:secSIM} we describe the simulation of ELG spectra using the Cosmos Mock Catalog as well as the expected redshift success rate when spectra are observed with the BOSS spectrograph. Section~\ref{escoffier:secOBS} concludes with the visual inspection of $2,000$ spectra observed during an ancillary program of BOSS dedicated to this study.  

%%-------------------------
\section{The ELG color selection}\label{escoffier:secELG}
\begin{figure}[ht!]
 \centering
 \includegraphics[width=0.4\textwidth,clip]{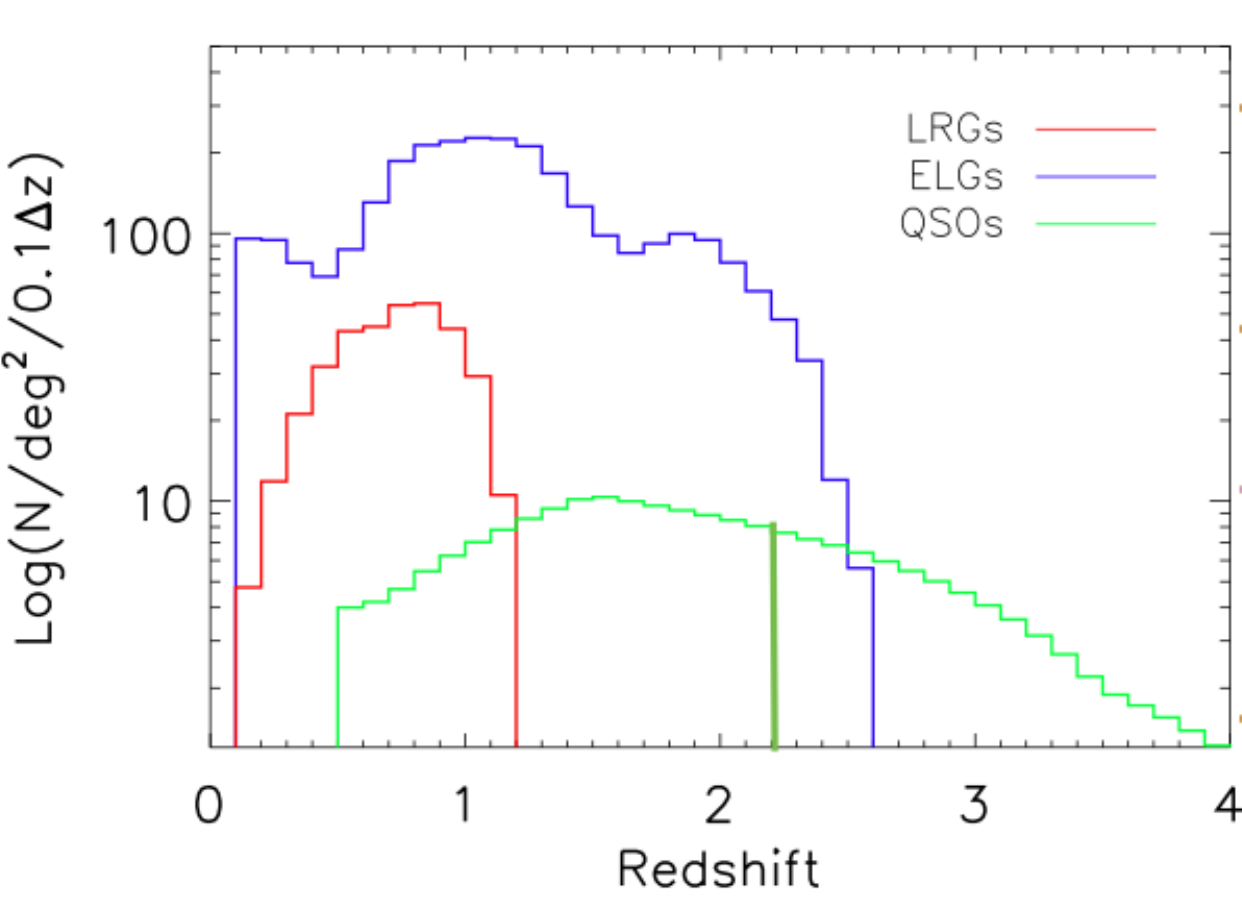}      
  \caption{n(z) distribution for targets selected for BigBOSS (courtesy of Nick Mostek, BigBOSS collaboration).}
  \label{escoffier:fig1}
\end{figure}

%%-------------------------
%\subsection{Motivation}
At lower redshift, luminous red galaxies are good candidates for spectroscopic surveys as they have strong absorption features like the $4000\AA$ break. However, from $z>1$, the $4000\AA$ break moves into the infrared and, as red galaxies become very faint, long exposure time is required. Moreover density of LRG targets falls dramatically at $z>1$ as shown in Fig.~\ref{escoffier:fig1}. At high redshift near $z\sim3$, a challenging option is to target quasars (QSOs). By instance, one of the key goals of the BOSS project is to study BAO features using Lyman-$\alpha$ forest absorption spectra of distant quasars in the range $2.2<z<3.5$ \citep{McDonald07,Ross12}. When working at $z\sim1$, a relevant choice is emission-line galaxies, including both strongly star-forming galaxies and emission-line QSOs. In this paper, we only consider star-forming emission-line galaxies that we call ELG. One of the advantages of ELG is that emission lines can be detected even when the continuum is weak. The redshift extraction is based on identification of emission lines, with the detection of the [OII] {$\lambda$}3727 doublet for $0.6<z<1.6$.

%\subsection{SDSS and CFHTLS photometry}
The aim of this work is to apply selection criteria to BOSS that uses the 2.5-m SDSS telescope at Apache Point Observatory \citep{Gunn06}. The SDSS photometric survey, delivered under the data release 8 (DR8, \citep{Aihara11}), covers $14,555$ square degrees in the 5 photometric bands $u, g, r, i, z$. The $3\sigma$ magnitude depths are: $u=22.0$, $g = 22.2$, $r = 22.2$, $i = 21.3$. The magnitudes we use are corrected from galactic extinction.
In addition, we use the CFHTLS photometric redshift catalog \citep{Ilbert06,Coupon09}. The CFHTLS covers $155$ square degrees in the $u,g,r,i,z$ bands, with transmission curves of filters slightly different from SDSS. The $3\sigma$ magnitude depths are: $u=25.3$, $g=25.5$, $r=24.8$, $i=24.5$, thus the CFHTLS photometry is much deeper than SDSS DR8, however the CFHTLS covers only a small fraction of the SDSS field of view. The photometric redshift accuracy is estimated to be $\sigma_z < 0.04\;(1+z)$ for $g<22.5$. 

%\subsection{The color selection}
In this paper we explore two color selections based on $ugr$ and $gri$ color-color diagrams.

\begin{figure}[ht!]
 \centering
 \includegraphics[width=0.8\textwidth,clip]{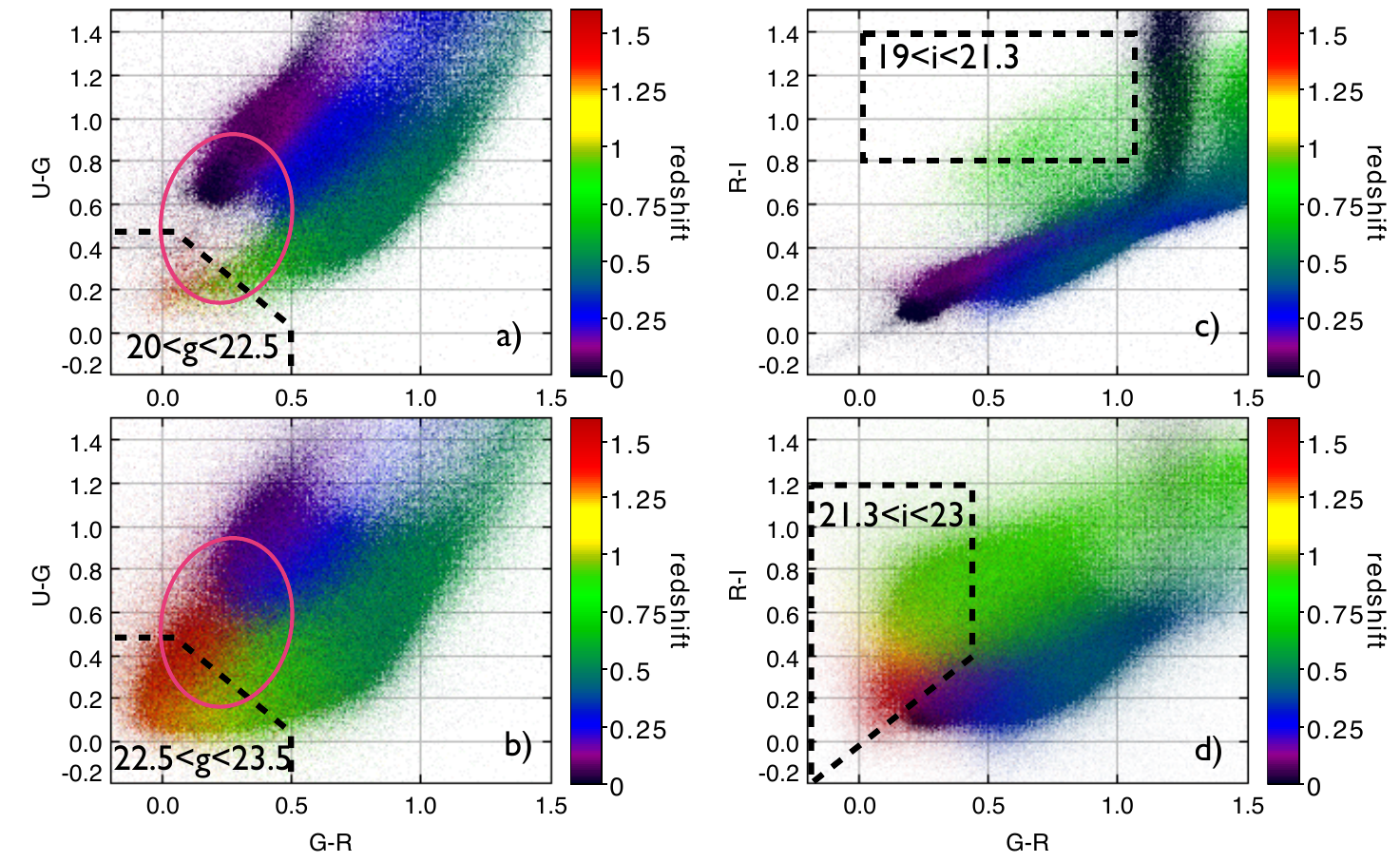}      
  \caption{Color-color diagrams in the CFHTLS photometric database. {\bf Left:} The $ugr$ selection for the bright (top) and the faint (bottom) samples. {\bf Right:} The $gri$ selection for the bright (top) and the faint (bottom) samples.}
  \label{escoffier:fig2}
\end{figure}

%\subsubsection{The ugr selection}
The $ugr$ color selection is defined by $-1<u-r<0.5$ and $-1<g-r<0.5$ that selects strongly star-forming galaxies at $z>0.6$. An additional cut $-1<u-g<0.5$ removes all low-redshift galaxies ($z<0.3$). For this selection, the bright sample is defined for the $g$ magnitude between $20<g<22.5$ and the faint sample for $g<23.5$ (Fig.~\ref{escoffier:fig2} Left). The $ugr$ color selection avoids the stellar sequence, but not the quasar sequence. Hence, the contamination of the $ugr$ selection by point-source objects is expected to be due to quasars. The selection is centered to $z\sim1.3$. 

%\subsubsection{The gri selection}
The bright sample of the $gri$ color selection is defined by the magnitude cut $19<i<21.3$, where in addition blue galaxies at $z\sim0.8$ are selected with $0.8<r-i<1.4$ and $-0.2<g-r<1.1$ (Fig.~\ref{escoffier:fig2} Right). In the faint range defined as $21.3<i<23$, we tilt the selection to select higher redshifts with $-0.4<g-r<0.4$, $-0.2<r-i<1.2$ and $g-r<r-i$. The $gri$ selection avoids both the stellar sequence and the quasar sequence. Thus the contamination from point-sources should be minimal. The selection is centered to $z\sim0.8$ (bright) and $z\sim1$ (faint). 

\section{Validation using simulations}\label{escoffier:secSIM}
%%-------------------------
%\subsection{Spectro-photometric catalogs}
In order to be able to evaluate the color selection efficiency, we need for a spectro-photometric catalog with a large sample of galaxy spectra. In this aim, we use the COSMOS Mock Catalog (CMC) based on the COSMOS photometric catalog \citep{Ilbert09}. The CMC contains about 280,000 galaxies, in which each simulated galaxy has a photo-z and a best-fit template \citep{Jouvel09}. In addition, with each galaxy of the CMC is associated a simulated spectrum, for which emission lines have been added using Kennicutt calibration laws \citep{Kennicutt98,Ilbert09} and calibrated using zCOSMOS \citep{Lilly09} as described in \citep{Zoubian12}. The strength of [OII] emission lines was confirmed using DEEP2 and VVDS DEEP luminosity functions \citep{Lefevre05,Zhu09}. Finally a host galaxy extinction law is applied to each spectrum.
%\subsection{Simulation and redshift extraction}
From the CMC, two simulated galaxy catalogs were built, one for each color selection function ($ugr$ and $gri$), for $0.6<z<1.7$. Each synthetic spectrum was affected by sky and photon noise as if observed by the BOSS spectrographs \citep{Smee12}, by using the {\it specsim1d} software. We simulated a set of four exposures of 900 seconds each. 

To extract the spectroscopic redshift, the resulting simulated spectra were then analyzed by the two redshift codes of the BOSS pipeline software, the idlspec2d software \citep{Bolton12} and a modified version of the Zcode Fortran program, initially developed by \citep{Sutherland99} and made available to the BOSS community. While the former, primarily designed for LRG targets, is based on a least-squares minimization using galaxies templates, the latter performed also a redshift estimate based on fitting discrete emission line templates in Fourier space over all z. In a further step we address the flux measurement of emission lines, conducted using the Platefit Vimos software developed by \citep{Lamareille09}. This software was developed to measure the flux of all emission lines after removing the stellar continuum and absorption lines from lower resolution and lower signal-to-noise ratio spectra \citep{Lamareille06}. 

\begin{figure}[ht!]
 \centering
 \includegraphics[width=0.5\textwidth,clip]{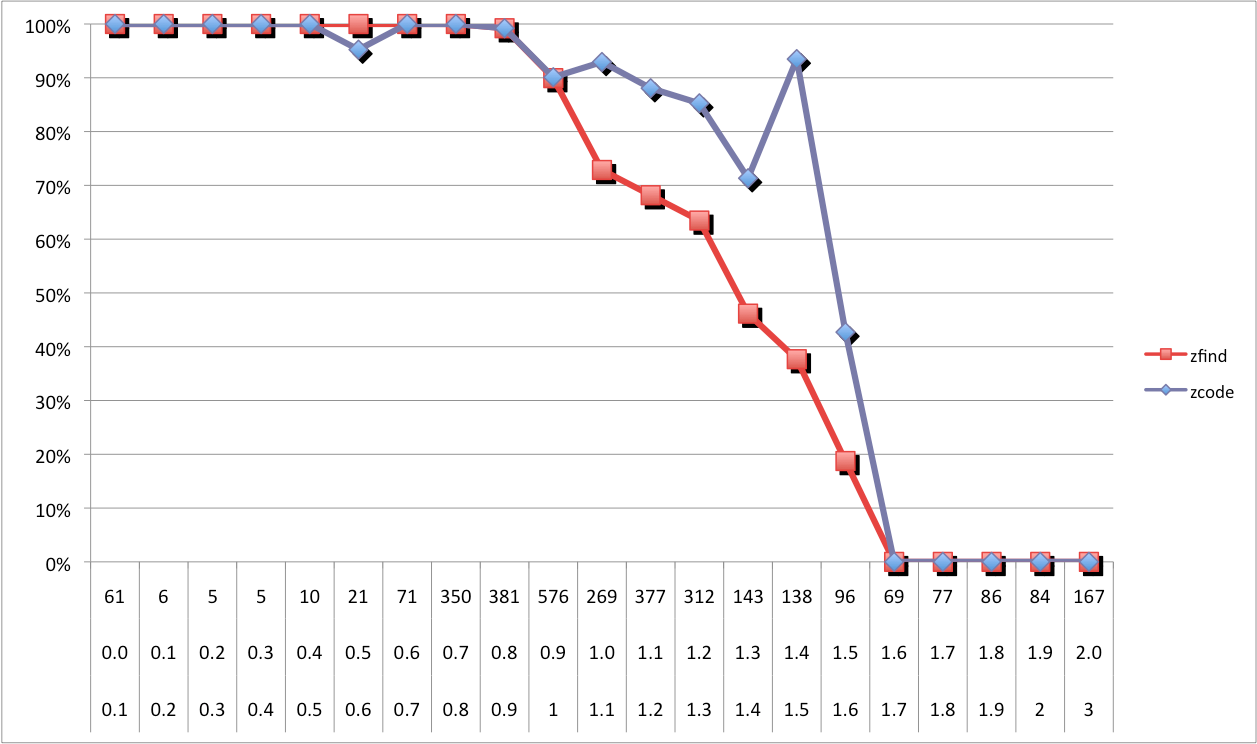}      
  \caption{Redshift success rate of the $urg$ sample with the two redshift codes of the BOSS pipeline.}
  \label{escoffier:fig3}
\end{figure}

Finally we define a successful redshift measurement if $\delta z / (1+z)<0.004$. The redshift success rate (RSR) of the $ugr$ selection is shown in Fig.~\ref{escoffier:fig3} for the two redshift codes of the BOSS pipeline, using galaxies templates (in red) and emission lines templates (in blue). The redshift extraction is not meaningful below $z<0.6$ due to the lack of statistics. The
RSR is higher than $70\%$ for $z<1.5$, and is about $90\%$ for the n(z) distribution between $0.6<z<1.6$. For $z>1.6$, the redshift extraction always fails, and some catastrophic failures ($\delta z / (1+z)>0.01$) remain for $1<z<1.6$, essentially due to the bad identification of the [OII] emission line view as the $H\alpha$ line.

%%---------------------

\section{Validation using ancillary observations}\label{escoffier:secOBS}
%\subsection{The BOSS ELG ancillary program}
To test the reliability of both the bright $ugr$ and $gri$ color selections, we have conducted a set of dedicated observations, as part of the Emission Line Galaxy SDSS-III/BOSS ancillary program. The observations were conducted between Autumn 2010 and Spring 2011 using the SDSS telescope with the BOSS spectrograph. A total of $\sim2,000$ spectra, observed four times of 15 minutes, were taken in different fields: in the Stripe 82 (using single epoch SDSS photometry for color selection) and in the CFHTLS W1, W3 and W4 wide fields (using CFHT-LS photometry). This data set has been released in the SDSS-III DR9 \citep{Ahn12}.

All these observed spectra were visually inspected to confirm or correct the redshifts produced by two different pipelines. To classify the observed objects, we have defined sub-categories, one with secure redshifts and the other with unreliable redshifts. For the targets selected using the SDSS photometry and the $ugr$ color selection : 32\% are ELGs at z > 0.6, and 32\% are ELGs at $z<0.6$. Besides, 20\% are flagged as bad data, and QSOs are 10\% of the selected targets. Within the $gri$ selection, 50\% of targets are ELGs at $z>0.6$. 

Using the CFHTLS photometry, 46\% of targets are ELGs at z > 0.6 and 14\% are QSOs with the $ugr$ selection. Within the $gri$ selection, 61\% are ELGs at $z > 0.6$, 12\% are red galaxies with a strong continuum, and only 1\% of targets are QSOs. For both bright and faint samples, 18\% of spectra are bad data (for a complete description, see \citet{Comparat12}.

%%-------------------------

%\subsection{Hyperlinks}
%
%Hyperlinks can be introduced as follows: \url{http://www.sf2a.asso.fr/}.
%
%\subsection{Figures}
%%%-------------------
%
%You can either use (Encapsulated) PostScript files (compile with \LaTeX) or else files in the PDF format (process with \texttt{pdflatex}), as shown in Fig.~\ref{author1:fig1}. Two figures can be joined together as shown in Fig.~\ref{author1:fig2}. In this case, as illustrated in Fig.~\ref{author1:fig2}, 
%the caption must follow this format (e.g. boldface fonts for the Left and Right items): \textbf{Left:} text
%of the caption for the left panel. \textbf{Right:} text of the caption for the figure on the right hand side. 
%
%%%
%%% Example of single figure
%%%
%\begin{figure}[ht!]
% \centering
% \includegraphics[width=0.8\textwidth,clip]{author_fig1}      
%%% Note the ABSENCE of the extension .pdf , .eps or .ps  !
%  \caption{Caption here}
%  \label{author1:fig1}
%\end{figure}
%
%%%
%%% Example of two figures side by side
%%%
%\begin{figure}[ht!]
% \centering
% \includegraphics[width=0.48\textwidth,clip]{author_fig1}%      
% \includegraphics[width=0.48\textwidth,clip]{author_fig2}      
%%% Note the ABSENCE of the extension .pdf , .eps or .ps  !
%  \caption{{\bf Left:} Caption of the left panel. {\bf Right:} Caption of the right panel. }
%  \label{author1:fig2}
%\end{figure}

\section{Conclusions}
%%--------------------
We present an efficient emission-line galaxy selection that could provide a sample from which the BAO feature could be measured in the 2-point correlation function at $z > 0.6$. Using such deeper photometric surveys and improved pipelines, it should be possible to probe BAO to $z = 1.2$ in the next 6 years, e.g. by the eBOSS experiment, and to $z = 1.7$ in the next 10 years, e.g. by PFS-SuMIRE or BigBOSS experiment.

% Optional acknowledgements
% -------------------------
\begin{acknowledgements}
Funding for SDSS-III has been provided by the Alfred P. Sloan Foundation, the Participating Institutions, the National Science Foundation, and the U.S. Department of Energy Office of Science. The SDSS-III web site is \url{http://www.sdss3.org/.}
Based on observations obtained with MegaPrime/MegaCam, a joint project of CFHT and CEA/DAPNIA, at the Canada-France-Hawaii Telescope (CFHT) which is operated by the National Research Council (NRC) of Canada, the Institut National des Science de lÕUnivers of the Centre National de la Recherche Scientifique (CNRS) of France, and the University of Hawaii. This work is based in part on data products produced at TERAPIX and the Canadian Astronomy Data Centre as part of the Canada-France-Hawaii Telescope Legacy Survey, a collaborative project of NRC and CNRS.
This work was supported by the ANR grant ANR-08-BLAN-0222.
\end{acknowledgements}

\bibliographystyle{aa}  % A&A bibliography style file (aa.bst)
%\bibliography{sf2a-template} % your references in file: Yourfile.bib
\bibliography{escoffier} % your references in file: Yourfile.bib

\end{document}